\documentclass[prd,twocolumn,superscriptaddress,showpacs,nofootinbib,
preprintnumbers]{revtex4}

\usepackage{amsmath}
\usepackage{amsfonts}
\usepackage{graphicx}
\usepackage{dcolumn}

\def\be{\begin{equation}}
\def\ee{\end{equation}}
\def\ba{\begin{eqnarray}}
\def\ea{\end{eqnarray}}
\def\bs{\begin{subequations}}
\def\es{\end{subequations}}

\usepackage{color}

\begin{document}

\title{Prospects of inflation with perturbed throat geometry}
\author{Amna Ali}
\affiliation{Centre of Theoretical Physics, Jamia Millia Islamia,
New Delhi-110025, India}

\author{R. Chingangbam}
\affiliation{Centre of Theoretical Physics, Jamia Millia Islamia,
New Delhi-110025, India}
\author{Sudhakar Panda}
\affiliation{Harish-Chandra Research Institute, Chhatnag Road,
Jhusi, Allahabad-211019, India}

\author{M.~Sami}
\affiliation{Centre of Theoretical Physics, Jamia Millia Islamia,
New Delhi-110025, India}

\begin{abstract}
We study brane inflation in a warped deformed conifold background
that includes general possible corrections to the throat geometry
sourced by coupling to the bulk of a compact Calabi-Yau space. We
focus specifically, on the perturbation by chiral operator of
dimension 3/2 in the CFT.  We find that the effective potential in
this case can give rise to required number of e-foldings and the
spectral index $n_S$ consistent with observation. The tensor to
scalar ratio of perturbations is generally very low in this
scenario. The COBE normalization, however,  poses certain
difficulties which can be circumvented provided model parameters are
properly fine tuned. We find the numerical values of parameters
which can give rise to enough inflation, observationally consistent
values of density perturbations, scalar to tensor ratio of
perturbations and the spectral index $n_S$.

\end{abstract}

\maketitle

\section{Introduction}
It is well known that the standard model for Big bang cosmology is
plagued with some intrinsic problems like horizon and flatness
problems. Such problems could be cured if one postulates that the
early universe went through a brief period of accelerated expansion,
otherwise called the cosmological inflation \cite{inflation}. The
inflationary scenario not only explains the large-scale homogeneity
of our universe but also proposes a mechanism, of quantum nature, to
generate the primordial inhomogeneities which is the seed for
understanding the structure formation in the universe. Such
inhomogeities have been observed as anisotropies in the temperature
of the cosmic microwave background, thus the theoretical ingredients
of inflation are subject to observational constraints. Moreover, the
paradigm of inflation has stood the test of theoretical and
observational challenges in the past two decades
\cite{Spergel1,Spergel2}. However, in spite of its cosmological
successes, it still remains a paradigm in search of a viable
theoretical model that can be embedded in a fundamental theory of
gravity . In this context, enormous amount of efforts are underway
to derive inflationary models from string theory, a consistent
quantum field theory around the Planck's scale. Progress in the
understanding of the nonperturbative dualities in string theory, to
be precise, discovery of D-branes, has further provided an important
framework to build and test inflationary models of cosmology.

In past few years, many inflationary models have been constructed in
the context of D-brane cosmology. It includes inflation due to
tachyon condensation on a non-BPS brane, inflation due to the motion
of a D3-brane towards an anti-D3-brane
\cite{sen,linde,kallosh,lindeD}, inflation due to geometric tachyon
arising from the motion of a probe brane in the background of a
stack of either NS5-branes or the dual D5-branes \cite{GTach} .
However, these models are based on effective field theory and assume
an underlying mechanism for the stabilization of various moduli
fields. Thus, they do not take into account the details of
compactification and the effects of moduli stabilization and hence
any predictions from these models are questionable.

Progress in search of a realistic inflationary model in string
theory was made when it was learnt that background fluxes can
stabilize all the complex structure moduli fields. In fact it has
been shown in Ref.~\cite{GKP} that the fluxes in a warped
compactification, using a Klebanov-Strassler (KS) throat \cite{KS},
can stabilize the axio-dilaton and the complex structure moduli of
type IIB string theory compactified on an orientifold of a
Calabi-Yau threefold. Further important progress was achieved when
it was shown in Ref.~\cite{KKLT} that the K\"ahler moduli fields
also can be stabilized by a combination of fluxes and
nonperturbative effects. The nonperturbative effects, in this
context, arise, via gauge dynamics of either an Euclidean D3-brane
or from a stack of   D7-branes wrapping super-symmetrically a
four cycle in the warped throat. The warped volume of the four cycle
controls the magnitude of the nonperturbative effect since it
affects the gauge coupling on the D7-branes wrapping this four
cycle.

Armed with these results, an inflationary model \cite{KKLMMT} has
been built taking into account of the compactification data (see
also Refs.~\cite{dbpapers}). The inflaton potential is obtained by
performing string theoretic computations involving the details of
the compactification scheme. In this setup inflation is realized by
the motion of a D3-brane towards a distant static anti-D3-brane,
placed at the tip of the throat and the radial separation between
the two is considered to be the inflaton field. The effect of the
moduli stabilization resulted in a mass term to the inflaton field
which is computed in \cite{KKLMMT}. It turned out, unfortunately
that the mass is large and of the order of hubble parameter and
hence spoils the inflation. To avoid this disappointment (see,
however, \cite{IT}), one needs to search for various sources of
correction to the inflaton potential. For example, in
Ref.~\cite{Bau1}, the embedding of the D7-branes as given in
\cite{Kup} was considered. Assuming that at least one of the
four-cycles carrying the nonperturbative effects descend down a
finite distance into the warped throat so that the brane is
constrained to move only inside the throat, it was found that such a
configuration leads to a perturbation to the warp factor affecting a
correction to the warped four cycle volume. Further, this correction
depends on the position of the D3-brane and thus the superpotential
for the nonperturbative effect gets corrected by an overall
position-dependent factor. Thus the full potential on the brane is
the sum of the potential (F-term) coming from the superpotential
and the usual D-term potential contributed by the interaction
between the D3-brane and the anti-D3-brane. Taking these corrections
into account, the volume modulus stabilization has been re-analyzed
in Refs.~\cite{Bau2,Bau3,KP} and  the viability of inflation was
investigated in this modified scenario. The stabilization of the
volume modulus puts severe constraint which is difficult to solve
analytically without invoking approximations. Moreover, the model
needs extreme fine tuning.

The above model has been reexamined in Ref.~\cite{PSS} where
inflation, involving both the volume modulus and the radial distance
between the brane and the anti-brane participate in the dynamics.
Making a rotation in the trajectory space, a linear combination of
the two fields, which become independent of time and hence can be
stabilized, is identified with the volume modulus. The orthogonal
combination then becomes the inflaton field. The model again needs
severe fine tuning and further it has been observed that when the
spectral index of scalar perturbation reaches the scale invariant
value, the amplitude tends to be larger than the COBE normalized
value by about three order of magnitude, making the model seemingly
unrealistic. However, a recent analysis using Monte Carlo method for
searching the parameter space shows that COBE normalization as well
as the requirement of nearly flat  spectrum can be satisfied at the
same time \cite{HC}.

In this paper, we analyse the possibility of a realistic model for
brane inflation remaining within the large volume compactification
scheme but incorporating a different correction to the inflaton
potential. Infact, the authors of Ref.~\cite{BDKKM} have recently
reported that there can be corrections to the inflaton potential
that arise from ultra violet deformations corresponding to
perturbation by the lowest-dimension operators in the dual conformal
field theory. These contributions to the potential have sensitive
dependence on the details of moduli stabilization and the gluing of
the throat into the compact Calabi-Yau space. In the next section we
briefly review the origin and form of these contributions to the
potential. In section 3, we analyse the inflationary dynamics based
on the full potential. Section 4 is devoted to the summery of our
analysis and conclusions.

\section{The D3-brane potential}

In this section we briefly outline the form of the scalar potential
on a mobile D3-brane in the set up of brane-antibrane inflation. The
fluxes for the compactification of type IIB string theory on an
orientifold of Calabi-Yau theory are chosen such that the internal
space has a warped throat region. An example of this background
geometry is the deformed conifold described in \cite{KS}. The form
of the potential for the inflaton field $\phi$ can be schematically
split into three parts; $V (\phi)~=~V_D (\phi)~+~V_F (\phi)~+~
V_\Delta (\phi)$ where $V_D$ is the combination of the warped
anti-D3-brane tension and the attractive Coulomb potential between
the D3 and anti-D3 branes. $V_F$ is the F-term potential, which
gives rise to a mass term, related to the moduli stabilization
effects. The exact form of $V_D$ and $V_F$ have been known for some
time now \cite{KKLMMT} and we do not elaborate it further. Rather,
we draw our attention to the other constituent of the potential,
namely $V_\Delta$ as proposed in \cite{BDKKM}. The authors, using
AdS/CFT duality, describe the compactified throat region as an
approximate conformal field theory which is cut off at some high
mass scale $M_{UV}$. The throat is taken to be long enough so that
the gauge theory is approximately conformal over a wide class of
energy scales. The mobile D3-brane is considered to be
well-separated from both the ultra violet and infrared regions.
Further, it has been assumed that all the moduli are stabilized
following Ref.\cite{KKLT}. In the present scenario, the authors
observe that there are bulk moduli fields $X$ with F-terms
$F_X~\sim~\xi a_0^2$ where $a_0$ is the minimal warp factor in the throat
and the field $X$ can also be thought of as an
open string modulus superfield with it lowest component being the
inflaton itself. The value of $\xi$ is fine-tunable depending on the
bulk fluxes, choice of sources for supersymmetry breaking or
Calabi-Yau geometries. These bulk moduli fields can couple to the
fields in the conformal field theory and this coupling leads to a
perturbation to the K\"ahler potential ${\cal K}$ of the form: \be
{\cal K}_\Delta~ =~c \int d^4\theta M_{UV}^{-\Delta} X^\dagger X
{\cal O}_\Delta \ee where $c$ is a constant and ${\cal O}_\Delta$ is
a gauge invariant operator of dimension $\Delta$ in the conformal
gauge theory dual to the throat. $M_{UV}$ is the scale which relates
to the UV cutoff of the gauge theory corresponding to the large
radial distance limit of the throat geometry. The above perturbation
yields a scalar potential of the form: \be V_\Delta~=~c
M_{UV}^{-\Delta} |F_X|^2 {\cal O}_\Delta . \ee The strategy followed
in \cite{BDKKM} is to find the operators $O_\Delta$ which are built
out of scalar fields that can give rise to a potential on the
Coulomb branch of the gauge theory and specially that contributes to
the potential which controls the radial motion corresponding to the
scalar field $\phi$. In this context, an interesting analysis  is
carried out to identify the leading non-normalizable modes
(normalizable in the full compact solution) of the supergravity
fields that can perturb the throat geometry, thus perturb the
D3-brane potential and arise from coupling to bulk moduli and four
dimensional supergravity. The Kaluza-Klein excitations around $AdS_5
\times T^{1,1}$ geometry \cite{Gub,CDDF,KRN} reveals that these
modes are linear combinations of the perturbations of the conformal
factor of $T^{1,1}$ and that of the four-form gauge field with all
four indices along $T^{1,1}$. From the harmonic expansions of these
modes, in the basis $SU(2) \times SU(2) \times U(1)_R$ global
symmetry, their radial dependence, $r^\Delta$ could be deduced and
moreover their contribution to the D3-brane potential could be
determined (see \cite{BDKKM} for details). It is observed that at
fixed radial location, the potential is minimized at some angular
location and when the brane sits at this angular location the radial
potential is negative. Further, the radial potential is minimized at
radial distance, $r \rightarrow \infty$. Writing the canonically
normalized inflation field as $\phi = \sqrt{T_3}r$, the potential is
found to be \be V_\Delta = - c a_0^4 T_3
(\frac{\phi}{\phi_{UV}})^\Delta \ee where $c \sim \cal{O}$ (1) is a
positive constant.

Taking into account the selection rules for the quantum numbers
associated with the global symmetry group, the leading corrections
to the inflaton potential comes from modes corresponding to $\Delta
= 2/3$ or $\Delta = 2$. The field theory dual to the warped deformed
conifold geometry is described by an $SU(N+M) \times SU(N)$ gauge
theory with bi-fundamental fields. One constructs single trace
operators involving these bi-fundamental fields and their complex
conjugates. These operators are also labeled by their $SU(2) \times
SU(2) \times U(1)_R$ quantum numbers. The dimensions of these
operators match with the dimensions of the perturbation modes
contributing to the inflaton potential which is verified using
AdS/CFT duality. It is observed that the operators having dimension
$\Delta = 3/2$ correspond to chiral operators and the the same for
$\Delta = 2$ correspond to non-chiral operators. Moreover, the
chiral operators determine the leading term in the inflaton
potential unless they are forbidden by symmetries which are
preserved in the string compactification. Infact, it is known that
these chiral operators are not present in the $Z_2$ orbifold of the
warped conifold. In such cases, the non-chiral operators determine
the leading contribution to the inflaton potential. Thus we have two
different models depending on the symmetries of the
compactification. We will focus on the details of inflation dynamics
for the model where there is no discrete symmetry forbidding the
chiral operators. For this case, combining with the $V_D$ and $V_F$
contributions, the full inflation potential is \be V(\phi)~=~D
\left[ 1 + \frac{1}{3} \left( \frac{\phi}{M_{pl}} \right)^2 -
C_{3/2} \left(\frac{\phi}{M_{UV}}\right)^{3/2}-\frac{3 D}{16 \pi^2
\phi^4} \right] \label{Spot} \ee

where $C_{3/2} \sim \cal{O}$(1) and $D\sim 2 a_0^4 T_3$ and $M_{UV}
\sim \phi_{UV}$.
We note that the functional form of the above potential is similar
to the single field form of the potential, after the volume modulus
field is fixed to the instantaneous minimum, of Ref.[18]. However,
the microscopic interpretation of the $\phi^{3/2}$ term is different
from the present context. This has been remarked in Ref.[22].
Moreover, the microscopic constraints on the coefficient of this
term in the potential plays a crucial role in the discussion of the
inflation dynamics. In fact, this constraint led to a two field
inflationary scenario of Ref.[20]. In the present model of Ref.[22],
these constraints no longer apply and hence it is necessary to
analyze the single field dynamics in the prevailing scenario of
relaxed constraints.

In the next section we shall discuss the background evolution of the
field $\phi$ with the potential (\ref{Spot}).


\section{Inflationary dynamics}
Let us investigate the possibilities of a viable inflation based
upon the potential (\ref{Spot}). In what follows it would be
convenient to cast the evolution equation for $\phi$ and the Hubble
equation
\begin{eqnarray}
&&\ddot{\phi}+3 H\dot{\phi}+V_{,\phi}=0   \\
&&H^2=\frac{1}{3 M_P^2}\left(\frac{\dot{\phi}^2}{2}+V(\phi)\right)
\end{eqnarray}
in the  autonomous form
\begin{eqnarray}
&&\frac{dx}{dN}=\frac{y}{{\cal H}} \\
&&\frac{dy}{dN}=-3y-\frac{1}{{\cal H}}\frac{ d{\cal V}}{dx}\\
&&{\cal H}^2=\frac{\alpha^2}{3}\left(\frac{1}{2}{y}^2+{\cal
V}(x)\right)
\end{eqnarray}
where $x=\phi/M_{UV}$, $y=\dot{x}/M_{UV}^{2}$, ${\cal H}=H/M_{UV}$,
${\cal V}=V/M_{UV}^4$ and $N$ designates the number of e-foldings.
The field potential is expressed through the dimensionless variable
$x$ as,
\begin{equation}
{\cal V}={\cal
D}\left(1+\frac{\alpha^2}{3}x^2-C_{3/2}x^{3/2}-\frac{3 {\cal D}}{16
\pi^2 \alpha^4 x^4}\right) \label{Spot2}
\end{equation}
where ${\cal D}=D/M_{UV}^4$, $\alpha=M_{UV}/M_P$. We should bear in
mind that $0< x < 1$ as $\phi< \phi_{UV} \sim M_{UV}$ and the mobile
D3-brane is moving towards the anti-D3 brane located at the tip of
the throat corresponding to $x=0$.
\begin{figure}
\begin{center}
\includegraphics[width=7cm,height=6cm,angle=0]{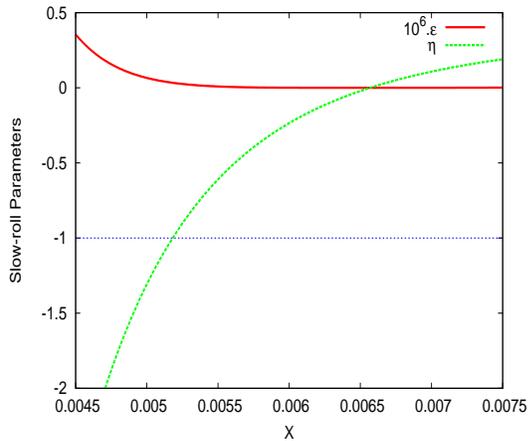}
\end{center}
 \caption{Plot of the slow roll parameters $\epsilon$ and $\eta$ for $\alpha^{-1}=2.000,~~ C_{3/2}=0.00982,~~ {\cal D}=1.000\times
10^{-15}$. The slow roll
 parameters are much smaller than one in the region of interest.  }
 \label{epeta}
\end{figure}
\begin{figure}
\begin{center}
\includegraphics[width=7cm,height=6cm,angle=0]{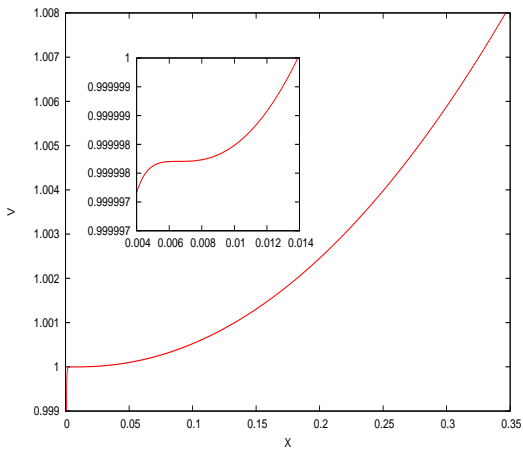}
\end{center}
 \caption{Plot of the effective potential for the same values of parameters as in Fig.\ref{epeta}. The insert shows the flat part of effective potential
near the origin which derives inflation.}
 \label{Pot1}
\end{figure}
\begin{figure}
\begin{center}
\includegraphics[width=7cm,height=6cm,angle=0]{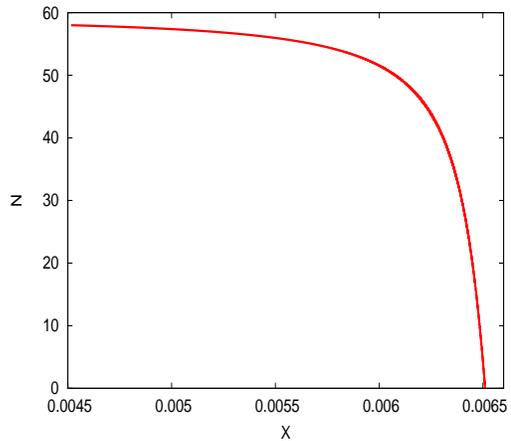}
\end{center}
 \caption{The number of e-folds $N$ versus the field $x$ from the beginning to the end of inflation. The model parameters are same as used in
 Fig.\ref{epeta}.   }
 \label{N1}
\end{figure}

 The slow roll parameters for the generic field range are
\begin{eqnarray}
&&\epsilon= \frac{1}{2 \alpha^2}\left(\frac{{\cal V}_{,x}}{{\cal
V}(x)}\right)^2 \simeq \frac{1}{2 \alpha^2}\Big[
\frac{2\alpha^2}{3}x-\frac{3 C_{3/2}}{2} x^{1/2} \nonumber \\
&&+\frac{3 {\cal D}}{4 \pi^2 \alpha^4 x^5}\Big ]^2 \\
&&\eta=\frac{1}{\alpha^2}\frac{{\cal V}_{,xx}}{{\cal V}(x)}\simeq
\frac{2}{3}-\frac{3C_{3/2}}{4\alpha^2}\frac{1}{x^{1/2}}-\frac{15{\cal
D}}{4 \pi^2 \alpha^6 x^6}\label{eta1}
\end{eqnarray}
Since $|\epsilon|<|\eta|$ in the present case, it is sufficient to
consider $\eta$ for discussing the slow roll conditions. It follows
from Eq.(\ref{eta1}) that $\eta$ is always less than one; it
decreases as $x$ moves towards the origin. At a particular value of
$x$, the slow roll parameter $\eta=-1$ marking the end of inflation
and takes large negative values thereafter for $x \to 0$, see
Fig.\ref{epeta}. For generic values of the model parameters, the
magnitude of slow roll parameter $\epsilon$ is much smaller than one
(see Fig.\ref{epeta}). Hence we have to worry here about $\eta$
alone.

\begin{figure}%
  \centering
  \parbox{3.1in}{\resizebox{85mm}{!}{\includegraphics{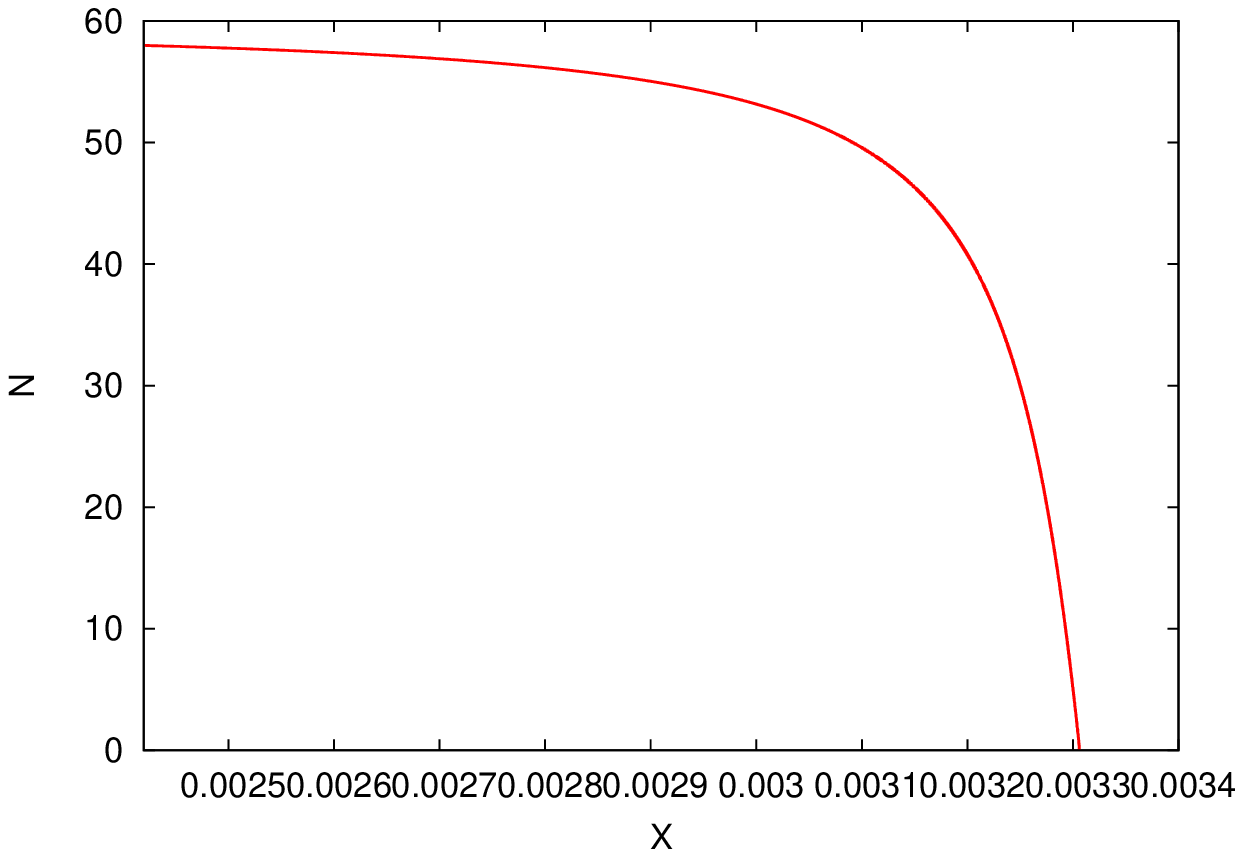}}
 \caption{The number of e-folds $N$ versus the field $x$
 (in case of ${\cal D}=1.210\times 10^{-17}$ ,$\alpha^{-1}=2.11991$,$C_{3/2}=0.0062284$) as inflaton moves from
 $x_{int}=0.003305$ towards the origin. Inflation ends around $x=x_{end}=0.00240$
 after $60$ e-folds. }%
\label{N2}}%
  \qquad
  \begin{minipage}{3.1in}%
   \resizebox{85mm}{!}{\includegraphics{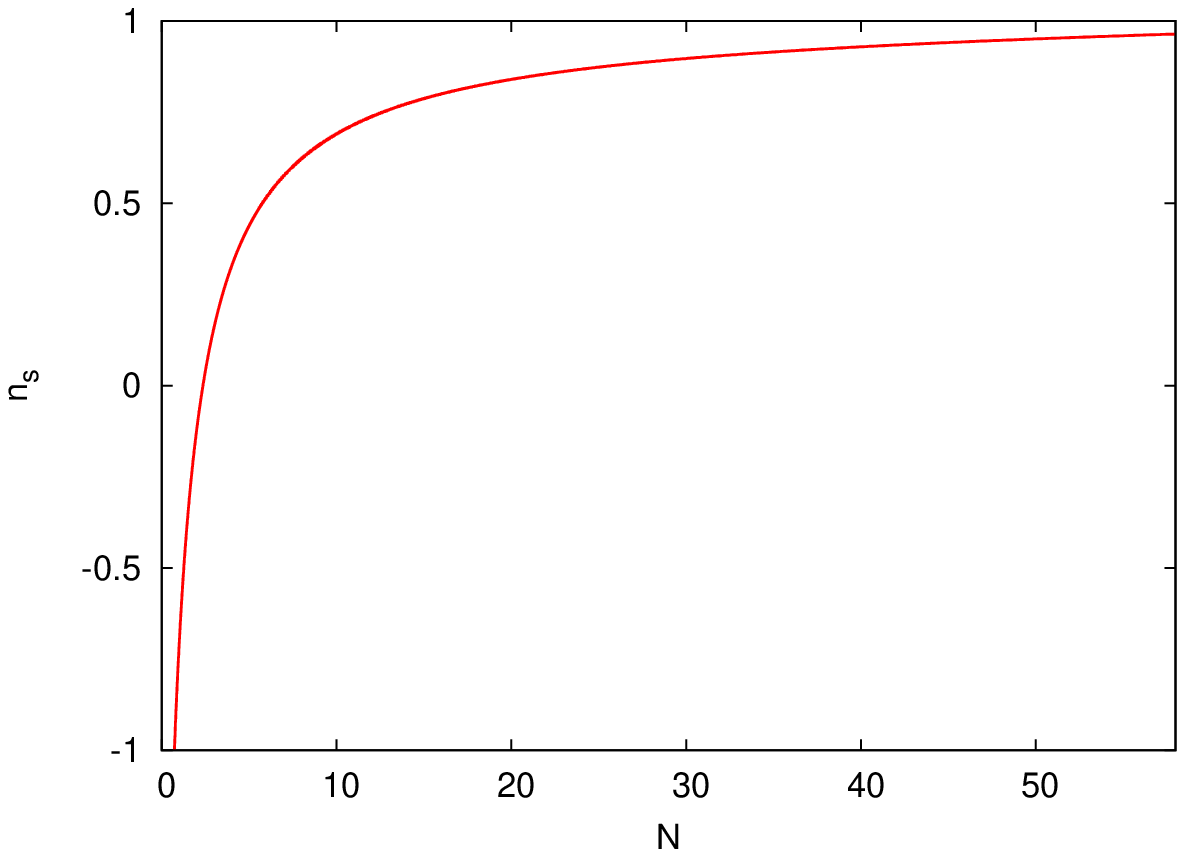}}
\caption{ Evolution of spectral index  $n_s$ versus the number of
e-folds, for the same parameters values as in Fig.\ref{N2}, starting
from the end of inflation. The spectral index
$ n_S \simeq 0.96$ for $N=60$.}%
\label{Sp2}%
  \end{minipage}%
\end{figure}

\begin{figure}
  \centering
  \parbox{3.1in}{\resizebox{85mm}{!}{\includegraphics{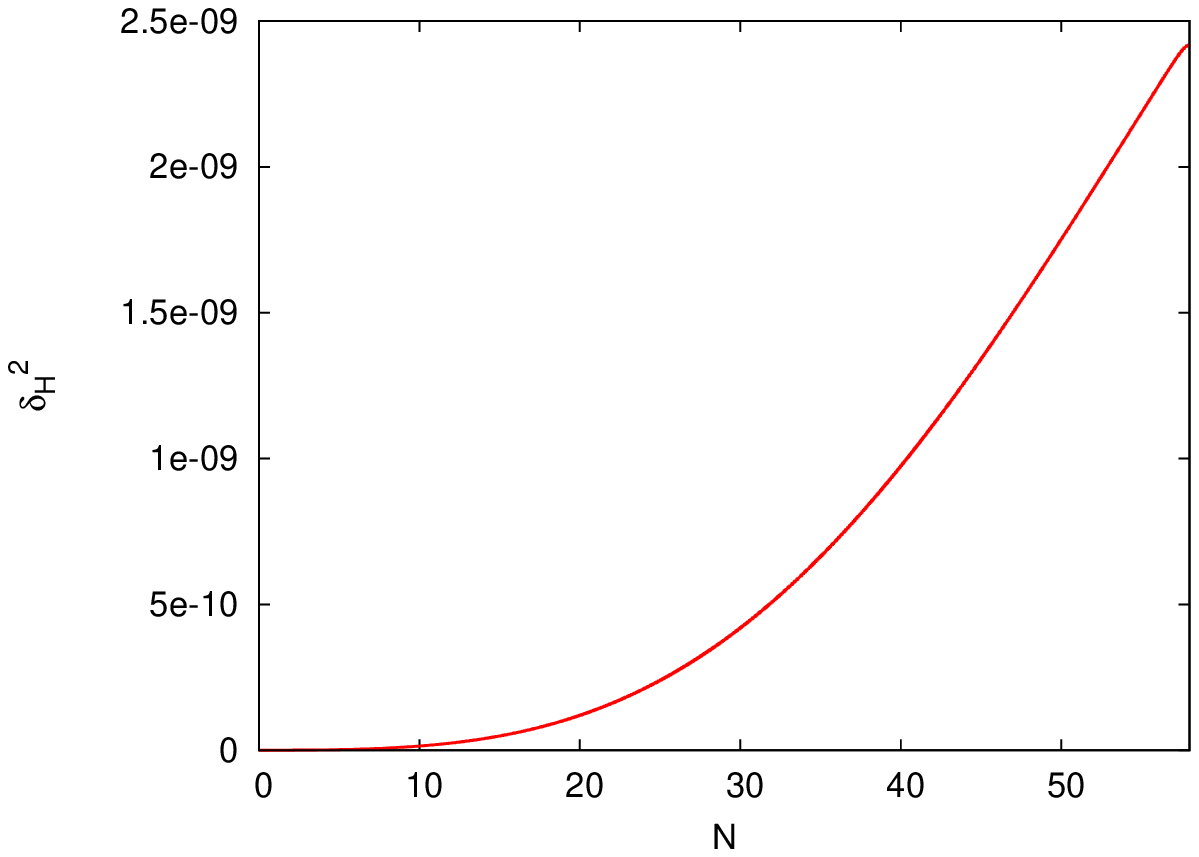}}
 \caption{Evolution of the amplitude of density perturbation $\delta_H^2$. versus $N$
 beginning from the end of inflation. $\delta_H^2\simeq 2.4\times 10^{-9}$ for $N=60$. The model parameters are same as in
 Fig.\ref{N2}}%
\label{Ps1}}%
  \qquad
   \resizebox{85mm}{!}{\includegraphics{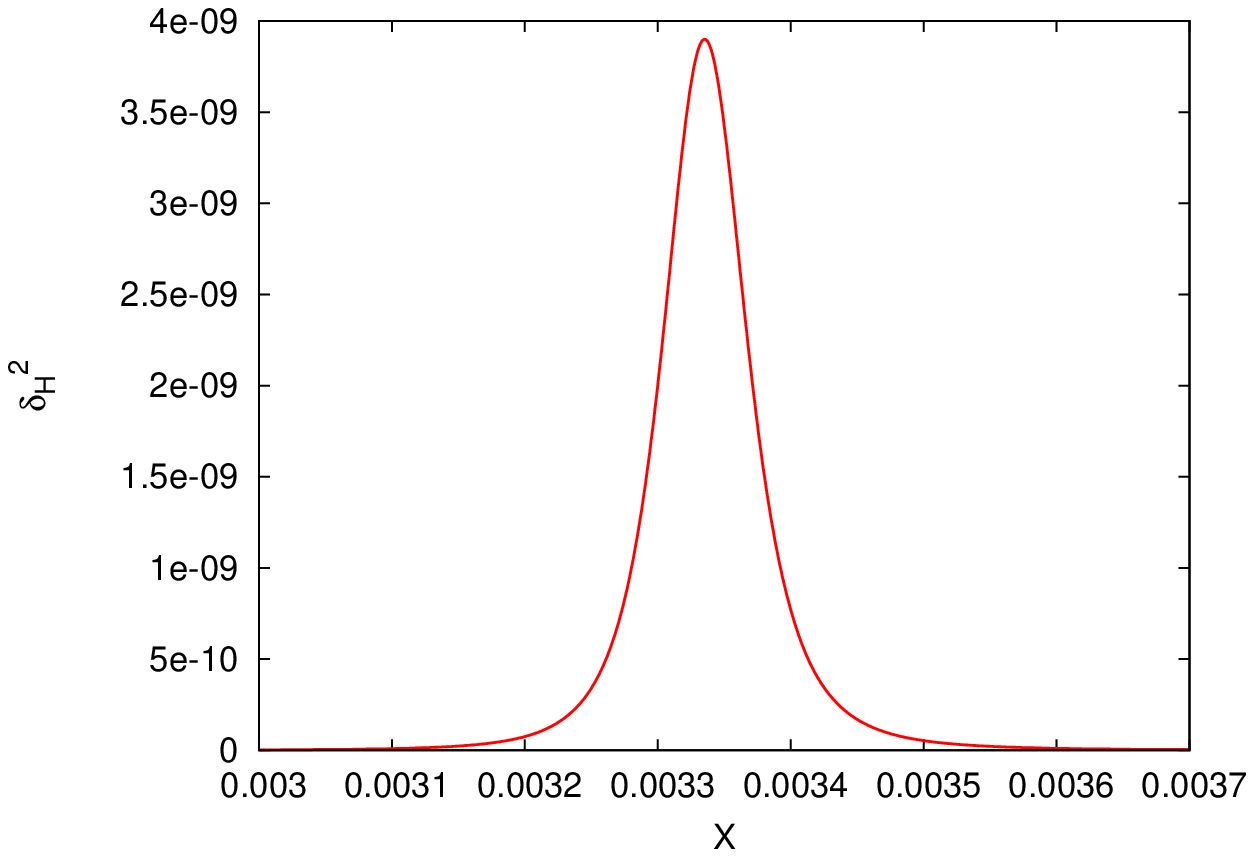}}
\caption{ Plot of $\delta_H^2$ versus the field $x$ for parameter
values same as in Fig.\ref{N2}. Inflation commences when
$x=x_{int}=0.003305$.} \label{Cobe1}
\end{figure}

In the case under investigation, the field $x$ rolls from $x=1$
towards the origin where $\Bar{D}3$ brane is located. Thus the field
potential should be monotonously increasing function of $x$.
Depending upon the numerical values of the model parameters, the
field potential (\ref{Spot2}) may be monotonously increasing
(decreasing) or even acquiring a minimum for $0<x<1$. The parameters
should therefore be chosen such that the potential has the right
behavior,
\begin{equation}
{\cal V}_x=\frac{2}{3}\alpha^2x^2-C_{3/2}x^{3/2}+\frac {3 {\cal
D}}{16 \pi^2 \alpha^4 x^5}>0 \label{Vx}
\end{equation}
Before we get to numerics, let us emphasize some general features of
the model. The last term in Eq.(\ref{Vx}) is always positive; hence
monotonicity of ${\cal V}(x)$ is ensured provided, $C_{3/2} \le 9
\alpha^2 x^{1/2}/4$ which imposes a constraint on the coefficient,
$C_{3/2}$.
 To
avoid minimum, we require smaller and smaller values of $C_{3/2}$ as
we move towards the origin before the last term in Eq.(\ref{Vx})
could take over. It turns out that ${\cal D}^{1/4} \sim 10^{-4}$ for
observational constraints to be satisfied pushing $C_{3/2}$ towards
numerical values much smaller than one. It is then possible to make
the potential flat near the origin, see Fig.\ref{Pot1}. Since the
field range viable for inflation is narrow, the potential should be
made sufficiently flat to derive required number of e-foldings. In
particular, it means that the sow roll parameter $\epsilon$ is very
small leading to low value of tensor to scalar ratio of
perturbations. This feature, however, becomes problematic for scalar
perturbations. Indeed, since $\delta_H^2 \propto {\cal V}/\epsilon$
and ${\cal V} \sim {\cal D}$, smaller values of $\epsilon$ lead to
larger values of density perturbations. Last but not the least, we
emphasize a peculiarity of the potential associated with the last
term in (\ref{Spot}). The constant ${\cal D}$ does not only appear
as an over all scale in the expression of the effective potential,
it also effects the behavior of $V(x)$ in a crucial manner, for
instance it changes the slow roll parameters, which also makes it
tedious to set the COBE normalization. If it were not so, it would
have been easy to satisfy the COBE constraint. This makes the search
difficult for viable parameters.
\begin{center}
\begin{figure}
\includegraphics[width=7cm,height=6cm,angle=0]{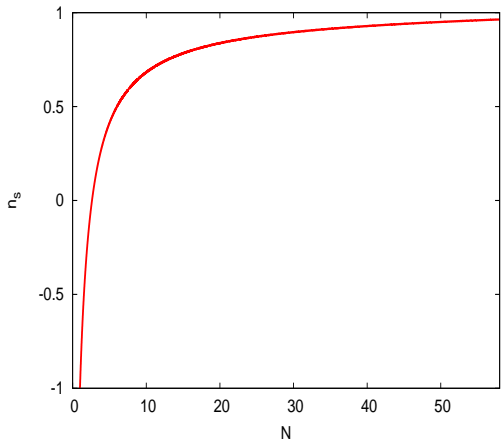}
 \caption{ Evolution of spectral index  $n_s$ versus the number of e-folds starting from the end of inflation.
 $n_S$ reaches the observed value after $60$ e-folds. } \label{Sp1}
\end{figure}
\end{center}
In what follows we discuss our numerical results. We have evolved
the equations of motion numerically by varying the model parameters
${\cal D}$,$ \alpha$ and $ C_{3/2}$. Since $\alpha \equiv M_P/M_{UV}
<1$ and $C_{3/2}<9 \alpha^2 x^{1/2}/4$ gives $C_{3/2}<9/16$ for
$\alpha=1/2$. Avoidance of local minimum of the effective potential
as the $D3$ brane moves towards the origin requires smaller and
smaller values of $C_{3/2}$ before the Coulomb part of potential
takes over. Since ${\cal D}$ does not quite define the scale of the
potential, it becomes necessary to vary all the parameters to meet
the observational constraints. It is possible to make the potential
flat near the origin and since the field range viable to inflation
is small for generic cases, the parameters should be suitably
adjusted allowing sufficient number of e-folds. In this case, it is
easy to obtain the required number of e-folds, flat power spectrum
and low value of tensor to scalar ratio of perturbations. However,
it is little tedious to get the COBE normalization,
\begin{equation}
\delta_H^2 \simeq \frac{1}{150 \pi^2M_P^4} \frac{V}{\epsilon}=
\frac{\alpha^4}{150 \pi^2}\frac{{\cal V}}{\epsilon}
\end{equation}
right as $\epsilon$ is small in this case. In case, ${\cal
D}=1.00\times10^{-15},\alpha^{-1}=2.000, C_{3/2}=0.00982$, we have
displayed the plot of effective potential in Fig.\ref{Pot1} which is
flat near the origin. The slow roll parameters are small, specially,
$\epsilon<<1$ in this case, see Fig.\ref{epeta}. With these
parameters, it is possible to get the required number of e-folds,
see Fig.\ref{N1}. The evolution of the spectral index from the end
of inflation is shown in Fig.\ref{Sp1}; $n_S$ reaches the observed
value $n_S \simeq 0.96$ at $60$ e-folds. Unfortunately, the
amplitude of density perturbations, $\delta_H^2 \simeq 10^{-8}$, is
large in this case. It is not possible to set the COBE normalization
by merely changing $\alpha$ and $C_{3/2}$; it is necessary to vary
${\cal D}$ which in turn requires variation of other two parameters
for obtaining the desired values of $N$ and $n_S$. We looked for
other choices of parameters to satisfy the COBE normalization taking
smaller values of ${\cal D}$ which further narrows the field range
of interest leading to still smaller values of $\epsilon$. We could
find parameters,
$$ {\cal D}=1.210\times10^{-17},\alpha^{-1}=2.11991,C_{3/2}=0.0062284$$
which can give rise to sufficient inflation, observed values $n_S$
and $\delta_H^2$. In case, we start evolving from
$x=x_{int}=0.003305$, we can easily generate $60$ e-foldings, and
observed value of the spectral index $n_S \simeq 0.96$ (see Figs.
\ref{N2} $\&$ \ref{Sp2}). As shown in Fig.\ref{Ps1}, we also achieve
the observed value of density perturbation, $\delta_H^2 \simeq 2.4
\times 10^{-9}$, at cosmologically relevant scales observed by COBE.
We note we used the exact expressions for the slow roll parameters
in numerical simulations.
\begin{center}
\begin{figure}
\includegraphics[width=7cm,height=6cm,angle=0]{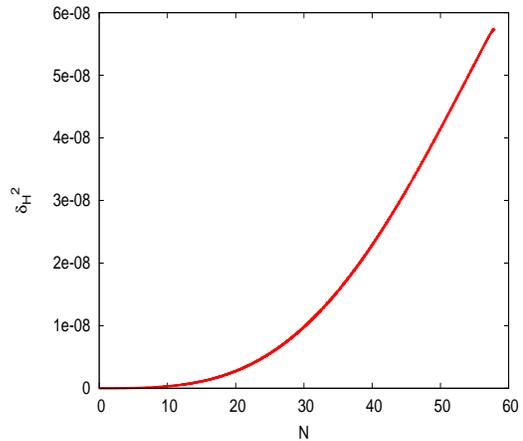}
 \caption{ Evolution of $\delta_H^2$ versus $N$ using the same parameters as in Fig.\ref{epeta}. $\delta_H^2$
 is larger than the observed values in this case.} \label{Power1}
\end{figure}
\end{center}
It is possible to change around the quoted values of parameters and
still satisfy the observational constraints. In Fig.\ref{Cobe1}, we
have plotted $\delta_H^2$ versus the field $x$ where inflaton rolls
towards the origin starting from $x=x_{int}= 0.003305$. Inflation
ends after $60$ e-folds around $x=x_{end}=0.0024$. Fig.\ref{Cobe1}
shows that $\delta_H^2 \simeq 2.4\times 10^{-9}$ at the commencement
of inflation. We note that $\delta_H^2$ peaks in the neighborhood of
$x=x_{int}= 0.003305$ and little deviation from the initial
condition can easily put the density perturbations out side its
observed limit. The parameters should be fine tuned to satisfy the
observational constraints, for instance, $C_{3/2}$ required to be
fine tuned to the level of one part in $10^{-7}$. The changes at the
seventh decimal puts physical quantities out side their
observational bounds or the potential can acquire local minimum and
spoil all the nice features of the model. Similarly other parameters
also need to be fine tuned. For instance, if we take ${\cal
D}=1.210\times 10^{-17}$ instead of ${\cal D}=1.211\times 10^{-17}$,
the field gets into the fast roll region before it could derive $60$
e-folds and COBE normalization is spoiled. The parameter
$\alpha^{-1}$ also requires fine tuning of the order of one part in
$10^{-5}$. It is quite possible that the systematic search of
parameters based on Monte-Carlo method discussed in Ref.\cite{HC}
might help to alleviate the fine tuning problem. It is,
nevertheless, remarkable that the correction to throat geometry
sourced by coupling to bulk allows not only to solve the well know
$\eta$ problem of D-brane cosmology but also helps in satisfying all
the observational constraints given by the WMAP5.

\section{Conclusions}
In this paper we have analysed the possibilities of inflation in a
warped background with an effective potential (\ref{Spot}). The
model includes three parameters, ${\cal D}$, $\alpha=M_{UV}/M_P$ and
$C_{3/2}$. The fact that $D3$ brane moves towards the tip of the
throat and can reach close to it and not vise-versa imposes
constraints on the model parameters, namely, $C_{3/2} \lesssim 0.1$
for a viable range of $\alpha$. The COBE normalization demands that
typically, ${\cal D}^{1/4}\sim 10^{-4}$ which makes $C_{3/2}$ much
smaller than one. Inflation becomes possible near the origin where
potential can be made sufficiently flat by appropriately choosing
the parameters. Numerical values of model parameters can easily be
set to obtain enough inflation and observationally consistent value
of $n_S$. For instance, in case of, $\alpha^{-1}=2.000,
C_{3/2}=0.00982,{\cal D}=1.000\times 10^{-15}$, we find that
$N\simeq 60$ and $n_S \simeq 0.96$. The tensor to scalar ratio is
very low in this case. The problem is caused by COBE normalization;
the amplitude of density perturbations is large in this case,
$\delta_H^2 \simeq 6\times 10^{-8}$ . This is related to the fact
that the constant ${\cal D}$ does not give the over all scale of the
potential but crucially effects the slow roll parameters and the
spectral index. Infact, for several other choices of ${\cal D}$, we
can obtain the required values of $N$ and $n_S$ keeping the low
value of tensor to scalar ratio of perturbations. However, it is
tricky to satisfy  the COBE normalization. Our search led to the
viable numerical values of the parameters and we have shown that for
${\cal D}=1.210\times
10^{-17},~~\alpha^{-1}=2.11991,~~C_{3/2}=0.0062284$; the total
number of e-foldings after horizon crossing  $N$ is around $60$ and
$n_S\simeq 0.96$ at the cosmologically relevant scales. For this
choice of parameters, the COBE normalization can be set correctly
and we find that $\delta_H^2 \simeq 2.4\times 10^{-9}$. It is really
interesting that in this case the tensor to scalar ratio is very
low, $r=16\epsilon \simeq 10^{-11}$, at the horizon crossing. It is
possible to vary the parameters around the given values and still
satisfy the observations constraints provided that we fine tune the
parameters ${\cal D}$, $\alpha$ and $C_{3/2}$. Our search in the
parameter space was carried out manually which allowed us to
demonstrate that the scenario under consideration can be made
consistent with the findings of WMAP5.

Finally, we should comment on the phenomenological aspects of model
discussed in Ref.[18] and compare it with the analysis presented
here. It was demonstrated in Ref.[20] that the single field models
based upon the scenario of Ref.[18] can give rise to only few
e-folds given the constrains on the model parameters. However, the
two field model (i.e. the volume modulus also participates in the
inflation dynamics) performs much better in this case. But it is
difficult to satisfy all the observational constraints
simultaneously [20], specially constraints of the spectral index and
the COBE normalization are problematic. In the scenario of
Ref.[22](where the microscopic theory is different from Ref.[18])
analyzed here, the constraints on model parameters are relaxed and
hence the phenomenology of single field inflation dynamics becomes
possible. For such a  single field model, we have demonstrated that
it is possible to satisfy all the observational constraints.



\section*{ACKNOWLEDGMENTS}
We are indebted to S. Kachru for a very important clarification. We
also thank I. Thangkool and T. Qureshi for valuable discussions. SP
thanks the Centre for theoretical physics, Jamia Millia University,
India and Centre for theoretical Physics, University of Groningen,
The Netherlands for their hospitality.


\begin{thebibliography}{40}

\bibitem{inflation} B.~A.~Bassett, S.~Tsujikawa and D.~Wands,
Rev.\ Mod.\ Phys.\  {\bf 78}, 537 (2006).



\bibitem{Spergel1} D.~N.~Spergel {\it et al.}, AstroAphys. J. Suppl. {\bf 148},
175(2003).

\bibitem{Spergel2} D.~N.~Spergel {\it et al.}, arXiv:astro-ph/0603452.

\bibitem{sen} A.~Sen, {\it Remarks on Tachyon Driven Cosmology};
arXiv:hep-th/0312153 and references therein.

\bibitem{linde} A.~D.~Linde, {\it Inflationary Cosmology}, arXiv:0705.0164 [hep-th].

\bibitem{kallosh} R.~Kallosh, {\it Inflation in string theory}, arXiv:hep-th/0702059.

\bibitem{lindeD}
G.~R.~Dvali and S.~H.~H.~Tye, Phys.\ Lett.\ B {\bf 450}, 72 (1999);
G.~R.~Dvali, Q.~Shafi and S.~Solganik, arXiv:hep-th/0105203;
C.~P.~Burgess {\it et al.}, JHEP {\bf 0107}, 047 (2001); G.~Shiu and
S.~H.~Tye, Phy. Lett. B {\bf 516}, 421(2001); S.~Alexander, \ Phys.
\ Rev.D {\bf 65}, 023507(2002); J.~Garcia-Bellido, R.~Rabadan and
F.~Zamora, JHEP {\bf 0201}, 036 (2002); M.~Gomez-Reino and
I.~Zavala,
JHEP {\bf 0209}, 020 (2002); C.~P.~Burgess, {\it et al.}, JHEP {\bf
0203}, 052(2002); N.~T.~Jones, H.~Stoica and S.~H.~H.~Tye, JHEP {\bf
0207}, 051 (2002); D.~Choudhury, D.~Ghoshal, D.~P.~Jatkar and
S.~Panda, JCAP {\bf 0307}, 009 (2003).

\bibitem{GTach} S.~Thomas and J.~Ward, \ Phys.\ Rev. D {\bf 72}, 083519(2005);
S.~Panda, M.~Sami, S.~Tsujikawa, \ Phys.\ Rev. D {\bf 73},
023515(2006); S.~Panda, M.~Sami, S.~Tsujikawa and J.~Ward, Phys.\
Rev.\  D {\bf 73}, 083512 (2006); B.~Gumjudpai, T.~Naskar and
J.~Ward, JCAP {\bf 0611}, 006 (2006); E.~Papantonopoulos, I.~Pappa
and V.~Zamarias, JHEP {\bf 0605}, 038 (2006); A. Sen,
arXiv:hep-th/0703157.

\bibitem{GKP} S.~B.~Giddings, S.~Kachru and J.~Polchinski,
Phys.\ Rev.\  D {\bf 66}, 106006 (2002).

\bibitem{KS} I.~R.~Klebanov and M.~J.~Strassler, JHEP {\bf 0008}, 052 (2000).

\bibitem{KKLT} S.~Kachru, R.~Kallosh, A.~Linde and S.~P.~Trivedi,
Phys.\ Rev.\  D {\bf 68}, 046005 (2003).

\bibitem{KKLMMT} S.~Kachru, R.~Kallosh, A.~Linde, J.~M.~Maldacena, L.~McAllister and
S.~P.~Trivedi,
JCAP {\bf 0310}, 013 (2003).

\bibitem{dbpapers} J.~P.~Hsu, R.~Kallosh and S.~Prokushkin, JCAP {\bf 0312}, 009
(2003); H.~Firouzjahi and S.~H.~H.~Tye, Phys.\ Lett.\  B {\bf 584},
147 (2004); C.~P.~Burgess, J.~M.~Cline, H.~Stoica and F.~Quevedo,
JHEP {\bf 0409}, 033 (2004); J.~J.~Blanco-Pillado {\it et al.}, JHEP
{\bf 0411}, 063 (2004); JHEP {\bf 0609}, 002 (2006); F.~Koyama,
Y.~Tachikawa and T.~Watari, Phys.\ Rev.\  D {\bf 69}, 106001 (2004);
J.~M.~Cline and H.~Stoica, Phys.\ Rev.\  D {\bf 72}, 126004 (2005);
R.~Allahverdi, K.~Enqvist, J.~Garcia-Bellido and A.~Mazumdar,
Phys.\ Rev.\ Lett.\  {\bf 97}, 191304 (2006); R.~Allahverdi {\it et
al.}, JCAP {\bf 0706}, 019 (2007).

\bibitem{IT} N.~Iizuka and S.~P.~Trivedi,
Phys.\ Rev.\  D {\bf 70}, 043519 (2004).

\bibitem{Bau1} D.~Baumann, A.~Dymarsky, I.~R.~Klebanov, J.~Maldacena, L.~McAllister
and A.~Murugan, JHEP {\bf 0611}, 031 (2006).

\bibitem{Kup} S.~Kuperstein, JHEP {\bf 0503}, 014 (2005).

\bibitem{Bau2} D.~Baumann, A.~Dymarsky, I.~R.~Klebanov, L.~McAllister and
P.~J.~Steinhardt,
arXiv:0705.3837 [hep-th].

\bibitem{Bau3} D.~Baumann, A.~Dymarsky, I.~R.~Klebanov and L.~McAllister,
arXiv:0706.0360 [hep-th].

\bibitem{KP} A.~Krause and E.~Pajer, arXiv:0705.4682 [hep-th].

\bibitem{PSS} S.~Panda, M.~Sami and S.~Tsujikawa, Phys.\ Rev.\ D {\bf 76} (2007)
103512.

\bibitem{HC} L.~Hoi and J.~Cline, arXiv:0810.1303 [hep-th].

\bibitem{BDKKM} D.~Baumann, A.~Dymarsky, S.~Kachru, I.~R.~Klebanov and
L.~McAllister, arXiv:hep-th/0808.2811.
\bibitem{Gub} S.~S.~Gubser, Phys.\ Rev.\ D {\bf 59} (1999) 025006.

\bibitem{CDDF} A.~Ceresole, G.~Dall'Agata, R.~D'Auria and S.~Ferrara, Phys.\ Rev. \
D {\bf 61} (2006) 066001.

\bibitem{KRN} H.~J.~Kim, L.~J.~Romans and P.~van Nieuwenhuizen, Phys.\ Rev.\ D
{\bf 32} (1985) 389.

\end{thebibliography}
\end{document}